\documentclass[seceq,supplement]{ptptex}

\usepackage{graphicx}
\usepackage{wrapft}

\newcommand{\beq}{\begin{eqnarray}}
\newcommand{\eeq}{\end{eqnarray}}

\newcommand{\del}{\partial}

\def\lsim{\displaystyle\mathop{<}_{\sim}}
\newcommand{\ket}{\rangle}
\newcommand{\bra}{\langle}

\newcommand{\dslash}{{\del \hspace{-6pt}/}}

\def\Journal#1#2#3#4{{#1} {\bf #2}, #3 (#4)}

\def\PRD{{\em Phys. Rev.} D}

\title{Chiral Symmetry Aspects of 
Positive and Negative Parity Baryons}

\author{
Atsushi \textsc{Hosaka}, Daisuke \textsc{Jido}~\footnote{
Present address:  ECT$^*$, Villa Tambosi, Strada delle 
Tabarelle 286, I38050 Villazzano (Trento), Italy}
and Makoto \textsc{Oka}$^{a}$%
}

\inst{
Research Center for Nuclear Physics,
        Osaka University, Ibaraki, Osaka 567-0047, Japan\\
$^a$ Department of Physics, 
Tokyo Institute of Technology, Tokyo 152-8551, Japan
}

\abst{
Chiral symmetry aspects for baryon properties are studied.  
After a brief discussion on general framework, we introduce 
two distinctive chiral group representations for baryons: 
the naive and mirror assignments.  
Using linear sigma models, nucleon properties are studied 
in both representations.  
Finally, we propose an experiment to distinguish the two assignments 
in the reactions of pion and eta productions.  
}

\begin{document}

\maketitle


%

\section{Introduction}

Chiral symmetry is one of fundamental symmetries of QCD and is 
considered to be very important in hadron physics.  
The fact that the observed particles do not fall into 
degenerate chiral multiplets (parity doublets) is an indication that 
the symmetry is broken spontaneously.  
Consequently, the almost massless Nambu-Goldstone bosons 
(pions and kaons) appear and their interactions are dictated by 
low energy theorems.  
Hadron properties are then governed by chiral symmetry with 
its spontaneous break down.  
In particular many of their properties are influenced by the 
pattern and strength of the symmetry breaking.  
Recent interests in the change in hadron properties at
finite baryonic densities and/or temperatures are also
related to the investigation of QCD phase properties.  

%

Particle spectrum is the most fundamental issue in order to 
obtain the information on realization of chiral symmetry.  
If chiral symmetry is manifest, 
the particle states are classified by irreducible representations of 
the chiral symmetry group.  
If not, however, the states are in general 
written as superposition of infinitely many terms of 
irreducible representations.  
Some time ago, by imposing suitable consistency conditions for 
scattering amplitudes of pions computed by using the low energy 
chiral lagrangian, Weinberg showed that observed hadrons may be classified 
into linear representations of the chiral symmetry 
group~\cite{weinberg1969,weinberg1990}.  
He also discussed  
examples of (1) $\pi, \sigma, \rho$ and $a_1$ 
and (2) $N$ and $\Delta$, where they 
form a larger representation of the chiral 
symmetry group~\cite{weinberg1990}.   
Such an investigation is interesting, since some observed  
quantities such as the axial 
coupling constants which can take any  number in  
the non-linear scheme can be determined.  
Recently investigations based on such chiral 
representation theories were also made in 
Refs.~\cite{jido4,beane1,beane2}.  
These algebraic methods are similar to the method of sum rule 
in the dispersion theory~\cite{adler1965}, but here the 
integrated states are saturated by one particle states.  
The method of chiral representations is also useful
for the discussion of property changes of hadrons 
when chiral symmetry is restored.  
In particular, the role of the sigma meson as a Higgs 
particle in the chiral theory can be studied in an 
explicit manner.  
The change in sigma meson properties at finite 
temperature and densities should be a clear indication 
of the importance of the chiral dynamics~\cite{kunihiro}.  

In this paper we discuss the role of chiral symmetry 
especially in the baryon sector.  
Besides the general remarks given in the above, we would like 
to clarify the meaning of the chiral representations for the 
positive and negative parity baryons.  
Since naively chiral symmetry transformations may relate states of 
opposite parities, it is natural to expect that some positive 
and negative parity states 
(parity doublet) form chiral representations.  
We will show that such identifications should be done carefully.  

First we discuss what chiral representations 
are possible for baryons.  
Among (in principle) infinitely many representations, we 
group them into two classes; one is what we call the naive 
assignment and the other the mirror 
assignment~\cite{jido2000,jido2001}.    
It is then shown that only in the mirror assignment, the positive and 
negative parity baryons fall into a single chiral representation.  
In contrast, in the naive assignment, 
the baryons with opposite parities are no longer 
related by chiral symmetry but are just independent particles.  
The two classes of chiral symmetry assignments for baryons may be 
distinguished by the sign of the axial coupling.  
We propose an experiment to observe the sign difference in  
pion and eta productions.

\section{Naive and mirror assignments of baryons}

Chiral symmetry is defined as a flavor symmetry for positive and 
negative chirality states.  
Here we consider for simplicity isospin symmetry.
Therefore, the chiral symmetry is expressed as 
$SU(2) \times SU(2)$.  
In QCD, it is for the (approximately) massless $u, d$ 
quark fields of 
two chiralities:
\beq
q_{r} = \frac{1 + \gamma_{5}}{2} q \, \; \; \; 
q_{l} = \frac{1 - \gamma_{5}}{2} q \, .  
\eeq 
In the massless limit, the chirality states are identified with 
the helicity states.  
Since hadrons are composite, their chiral representations can 
be, in principle, any one as labeled by $(p, q)$, 
where $p$ and $q$ are isospin values of the $SU(2)$ representations.  
In this notation, the quark field can be expressed as 
\beq
q_{r} \sim (1/2, 0)\, , \; \; \; q_{l} \sim (0, 1/2)\, .
\eeq
Due to parity invariance, if $h_{r}$ 
(a hadron with positive $=$ right-handed chirality)
$\sim (p, q)$, then
$h_{l} \sim (q, p)$, and physical states contain their 
equal weighted superposition 
$h = h_{r} \pm h_{l}$.  
Moreover, when spontaneous breaking occurs, the states are 
superposition of infinitely many terms, 
$
\sum_{pq} c_{pq} \, (p,q).  
$
According to the Weinberg's argument~\cite{weinberg1969}, 
however, there is a good reasoning to 
believe that hadrons belong to simpler (low dimensional) 
representations.  
In the following discussions, we consider such situations.  

Let us consider two nucleons, $N_{1}$ and $N_{2}$, where 
we assume that $N_{1}$ and $N_{2}$ carry 
positive and negative parities.  
Naively, we expect that they both belong to the same 
chiral representation; 
\beq
N_{1r}, N_{2r} \sim (p, q)\, , \; \; \; 
N_{1l}, N_{2l} \sim (q, p) \, .  
\label{naive}
\eeq
However, as pointed out previously~\cite{BWLee,weinberg1990}, 
it is also possible to consider an assignment as 
\beq
N_{1r}, N_{2l} \sim (p, q)\, , \; \; \; 
N_{1l}, N_{2r} \sim (q, p) \, ,
\label{mirror}
\eeq
where for the second nucleon the role of the right and left handed
components are interchanged.  
We call the first one of (\ref{naive}) naive assignment and the second 
one of (\ref{mirror}) mirror assignment.  

There are two significant differences in the two assignments.  
First, the sign of the axial coupling $g_{A}$ differs 
for $N_1$ and $N_2$. 
In the naive assignment, both $N_{1}$ and $N_{2}$ carry the same 
axial coupling, while in the mirror assignment, $g_A$ of $N_{2}$
has opposite sign of that of $N_{1}$.  
Second point concerns the mass parameters.  
In general two types are possible~\cite{weinberg1969}; 
one behaves as the 0-th component of a chiral 4-vector and the other 
behaves as a chiral scaler.  
The mass parameter of  chiral 4-vector may be related to the scale of 
chiral symmetry breaking $\bra \sigma \ket \equiv \sigma_{0}$, 
while the chiral scalar 
mass parameter could be independent of it.  
In the naive assignment, only mass terms of chiral 4-vector are allowed, 
while in 
the mirror assignment mass terms of both the chiral 4-vector and 
chiral scalar are allowed.

\section{Linear sigma models}

Let us consider linear sigma models for the naive and mirror 
models.  
For simplicity we consider the nucleon of the fundamental 
representations;  $N \sim (1/2, 0) + (0, 1/2)$.  
Furthermore, $\sigma$ and $\pi$ meson fields are introduced as
in the representation $(1/2, 1/2)$.  
Their transformation rules are then given by 
\beq
N_r  \to g_R N_r \, , \; \; \; 
N_l  \to g_L N_l
\eeq
and 
\beq
\sigma + i \vec \tau \cdot \vec \pi \to g_L 
(\sigma + i \vec \tau \cdot \vec \pi) g_R^\dagger \, .
\eeq 
where $(g_{R}, g_{L})$ are elements of 
$SU(2) \times SU(2)$.  

\subsection{Naive model} 

In the naive assignment, the chiral invariant 
lagrangian up to order (mass)$^4$ 
is given by~\cite{jido2000,jido2001}: 
\beq
L_{\rm{naive}} 
& = & 
\bar{N_1} i \dslash N_1 
- g_{1} \bar{N_1} 
(\sigma + i \gamma_5 \vec{\tau} \cdot \vec{\pi}) N_1 
+ \bar{N_2} i \dslash N_2 
- g_{2} \bar{N_2} 
(\sigma + i \gamma_5 \vec{\tau} \cdot \vec{\pi}) N_2 
\nonumber \\
& & 
- \ g_{12} \{ 
\bar{N_1} 
(\gamma_5 \sigma + i \vec{\tau} \cdot \vec{\pi}) N_2 
- 
\bar{N_2} 
(\gamma_5 \sigma + i \vec{\tau} \cdot \vec{\pi}) N_1  
\} 
+ {L}_{\rm mes}  \, , 
   \label{ordsu2lag}
\eeq
where the coupling constants 
$g_{1}$, $g_{2}$ and $g_{12}$ are free parameters.  
The terms of $g_{1}$ and $g_{2}$ are ordinary chiral invariant 
coupling terms of the linear sigma model.  
The term of $g_{12}$ is the mixing of $N_{1}$ and $N_{2}$.  
Since the two nucleons have opposite parity, $\gamma_{5}$ appears 
in the coupling with $\sigma$, while it does not in the coupling with 
$\pi$.  
The meson lagrangian ${L}_{\rm mes}$ in (\ref{ordsu2lag}) is 
not important in the following discussion.  

Chiral symmetry breaks down spontaneously when the sigma meson 
acquires a finite vacuum expectation value, 
$\sigma_{0} \equiv \bra 0 |\sigma |0 \ket$.  
This generates masses of the nucleons as a chiral 4-vector.  
From (\ref{ordsu2lag}), 
the mass term can be expressed by a $2 \times 2$ matrix in the two nucleon 
space of $N_{1}$ and $N_{2}$.   
The mass matrix can be diagonalized by the rotated states, 
\beq
    \left(
    \begin{array}{c}
	N_{+}  \\
	N_{-}
    \end{array}
    \right) 
    = 
    \left(
    \begin{array}{cc}
	\cos 2 \theta      & \gamma_{5} \sin 2 \theta  \\
	- \gamma_{5} \sin 2 \theta &  -\cos 2 \theta
    \end{array}
    \right)
    \left(
    \begin{array}{c}
	N_{1}  \\
	N_{2}
    \end{array}
    \right)  \, , \label{eigenN_nai}
\eeq
where the mixing angle and mass eigenvalues are given by 
\beq
\tan 2 \theta = \frac{2g_{1}}{g_{1} + g_{2}}\, , 
\; \; \; 
m_{\pm} = 
\frac{\sigma_{0}}{2} 
\left(
\sqrt{ (g_{1} + g_{2})^2 + 4g_{12}^2 }
\pm (g_{1} - g_{2} ) 
\right) \, . \label{anglemass1}
\eeq
In the naive model, since the interaction and mass matrices take the 
same form in the space of $2 \times 2$, 
the physical states, $N_{+}$ and $N_{-}$, decouple completely; 
the lagrangian becomes a sum of the $N_{+}$ and $N_{-}$ 
parts.  
Because of this, the coupling between the positive and negative parity baryons
disappears, 
which was considered as a part of the reasons for the small 
coupling $g_{N(1535) \to \pi N} \lsim 1$~\cite{jidoPRL}.  

Thus, chiral symmetry does not relate $N_{+}$ and $N_{-}$ in the 
naive model.  
The role of chiral symmetry in the naive model is nothing special.  
When chiral symmetry is restored and
$\sigma_{0} \rightarrow 0$,
both $N_+$ and $N_-$ become massless and degenerate.  
However, the degeneracy is trivial as they are independent.

\subsection{Mirror (Chiral doublet) model}

Let us turn to the mirror assignment.  
The lagrangian is given as~\cite{jido2000,jido2001}
\begin{eqnarray}
{L}_{\rm{mirror}} & = & 
\bar{N_1} i \dslash N_1 
- g_{1} \bar{N_{1}} 
(\sigma + i \gamma_5 \vec{\tau} \cdot \vec{\pi}) N_{1} 
+
\bar{N_2} i \dslash N_2
- g_{2} \bar{N_{2}} 
(\sigma - i \gamma_5 \vec{\tau} \cdot \vec{\pi}) N_{2} 
\nonumber \\
&-& m_{0}( \bar{N_1} \gamma_{5} N_2 - \bar{N_2} \gamma_{5} N_1  )
+ {L}_{\rm mes} \ .  
    \label{mirsu2lag}
\end{eqnarray}
Here we can introduce a mass term through the coupling between 
$N_1$ and $N_2$.  
We can verify that due to the transformation rule associated to 
(\ref{mirror}), this term is invariant under chiral symmetry 
transformations.  
Therefore, the mass parameter $m_0$ behave as a chiral scalar.  
This lagrangian (\ref{mirsu2lag}) 
was first formulated by DeTar and 
Kunihiro~\cite{DeTKun}.   

The mass matrix of the 
lagrangian (\ref{mirsu2lag}) 
can be diagonalized as in the naive model 
by a linear transformation.  
The mixing angle and mass eigenvalues are given by 
\beq
\tan 2 \theta = \frac{2m_{0}}{\sigma_{0}(g_{1} + g_{2})} \, , 
\; \; \; 
m_{\pm} =
\frac{\sigma_{0}}{2}
\left(
\sqrt{ (g_{1} + g_{2})^2  + 4\mu^2 }
\pm (g_{1} - g_{2} )
\right) \, ,
\label{anglemass2}
\eeq
where $\mu = m_{0}/\sigma_{0}$.  
Physical masses are written in terms of the two parameters; 
the chiral 4-vector $\sigma_{0}$ and the chiral scalar $m_{0}$.  
The nucleon masses can 
take a finite value $m_{0}$ when chiral symmetry is 
restored ($\sigma_{0} = 0$).  

The axial couplings can be computed from the 
commutation relations between 
the axial charge operators $Q_{5}^{a}$ and the 
nucleon fields.  
They take a matrix form as 
\beq
g_{A}= \left(
\begin{array}{c c}
    \cos 2\theta & -\sin 2\theta \\
    -\sin 2\theta & - \cos 2\theta
\end{array}
\right) \, .
\label{gAmirror}
\eeq  
From this we see that the signs of the diagonal components   
$g_{A}^{++}$ and $g_{A}^{--}$ have  opposite signs as expected.  
The absolute value is, however, smaller than one in contradiction with 
experiment, $g_{A} \sim 1.25$.  
Since the physical states $N_{\pm}$ are the 
superposition of $N_{1}$ and $N_2$, whose axial couplings are $\pm 1$, 
it follows that $|g_{A}^{++}|, |g_{A}^{--}| < 1$.  
If we consider a mixing with higher representation such as 
$(1, 1/2)$, the $g_{A}$ value can be larger~\cite{weinberg1969}.

It is interesting to see how the mirror model becomes consistent 
with experimental data.  
For this purpose, let us consider $N(939)$ and $N(1535)$ for the 
two nucleons $N_{+}$ and $N_-$.  
In order to fix the four
parameters, we use the inputs $m_{+} = 939, \; m_{-} = 1535$ MeV,
$\sigma_{0} = f_{\pi} = 93$ MeV and $g_{\pi N_{+} N_{-}} \sim
0.7$. 
The last relation
is extracted from the partial decay width of
$N(1535)$~\cite{PDG} (although large uncertainties for the width
have been reported~\cite{ManSal,VrDtLe}), 
\beq 
\Gamma_{N^* \to \pi
N} \sim 75 \; {\rm MeV} \, , 
\eeq
and the formula 
\beq 
\Gamma_{N^*
\to \pi N} = \frac{3q(E_{N} + M)}{4\pi M^*} g_{\pi NN^*}^2 \, ,
\eeq 
where $q$ is the relative momentum in the final state. 
We find 
\beq 
\sigma_{0} &=& 93 \, {\rm MeV}\, , \; \; \;
m_{0} = 270 \, {\rm MeV}\, , \; \; \; \nonumber \\
\label{paramfit} g_{1} &=& 9.8 \, , \; \; \; g_{2} = 16\, . 
\eeq
From these parameters, we obtain the mixing angle
\beq 
\theta = 6.3 ^{\circ} \, , 
\eeq 
giving the diagonal
value of the axial charges 
\beq 
\label{gApgAm} g_{A}^{++} = -
g_{A}^{--} = 0.98 \, . 
\eeq
Hence the mixing of the two states $N_{1,2}$ are not very large; 
in this scheme the nucleon $N(939)$ is dominated by the 
ordinary chiral component $N_1$, while $N(1535)$ by the mirror
component $N_2$.  

\subsection{Properties under chiral symmetry restoration}
 
\begin{figure}
       \centerline{\includegraphics[width=12cm]
                                   {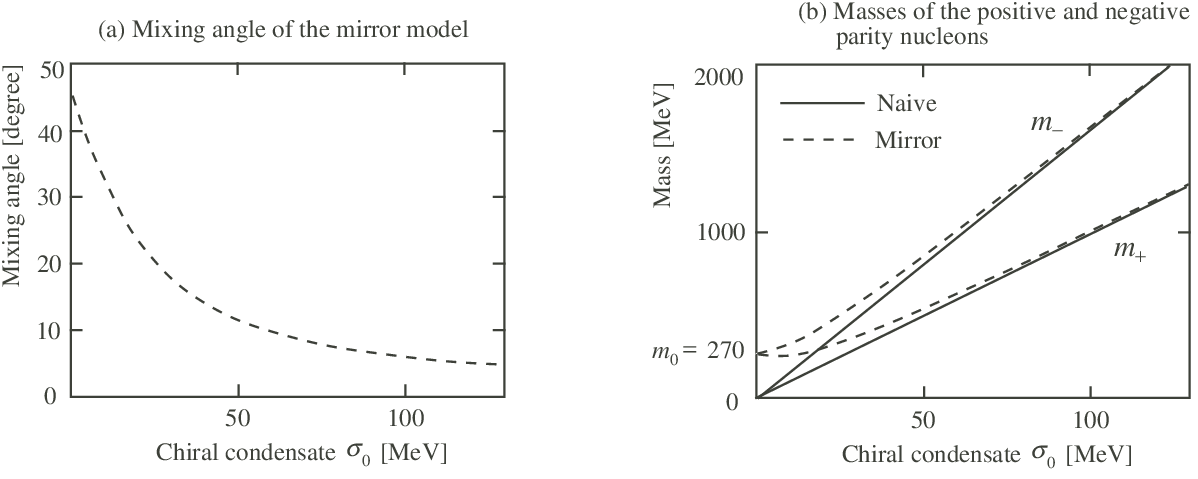}}
   \caption{(a) Mixing angle of the mirror model, and
   (b) the masses of the positive and negative parity nucleons in the
   naive and mirror models, as functions of the chiral condensate
   $\sigma_{0}$.}
   \label{mxangmass}
\end{figure}

\begin{figure}
       \centerline{\includegraphics[width=12cm]
                                   {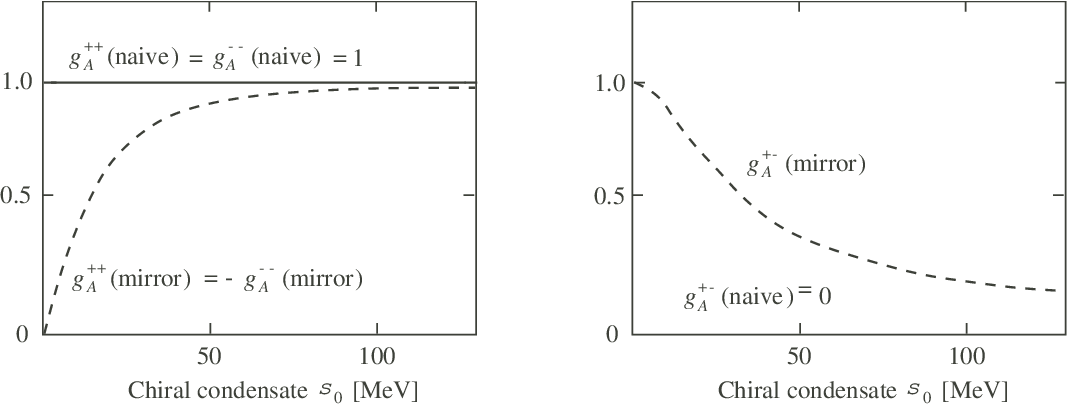}}
   \caption{Diagonal and off-diagonal axial charges in
   the naive and mirror models, as functions of the chiral condensate
   $\sigma_{0}$.}
   \label{plotga}
\end{figure}

The change in chiral symmetry is expressed  by the order parameter
$\sigma_{0}$, which is identified with the pion decay constant
$f_{\pi}$. It decreases (increases) as temperature or density 
increases (decreases).  
As $f_\pi$ varies, physical quantities receive the following 
changes:

 
\begin{itemize}
    \item Mixing angles and masses (Fig.~\ref{mxangmass}):\\
    When chiral symmetry is restored ($\sigma_{0} = 0$), the only
    source
    of mixing is the off-diagonal mass term in (\ref{mirsu2lag}).
    Therefore, the two degenerate states $N_{1}$ and $N_{2}$
    mix with equal weight ($\theta = 45 ^{\circ}$), having equal masses
    $m_{\pm} = m_{0}$.
    As the interaction is turned on and chiral symmetry starts to
    break spontaneously, the mixing angle decreases monotonically and
    reaches $6.3 ^\circ $ in the real world ($\sigma_{0} = 93$ MeV).
    Also, as $\sigma_{0}$ is increased,
    the masses increase, with the degeneracy being resolved.  
    Eventually, masses take their experimental values
    $m_{+} = 939$ MeV and $m_{-} = 1535$ MeV when
    $\sigma_{0} = 93$ MeV.
 
    \item $g_{A}$ (Fig.~\ref{plotga})\\
    It is interesting to see the behavior of the diagonal
    $g_{A}^{++} = -g_{A}^{--}$ and off-diagonal
    $g_{A}^{+-}$ as functions of $\sigma_{0}$.
    In particular, $g_{A}^{+-}$ increases as chiral symmetry
    begins to be restored.
    Since the axial charges are related to pion couplings, the
    increase of the off-diagonal couplings
    $g_{A}^{+-} \sim g_{\pi N_{+} N_{-}}$ has been considered as
    a cause of the increase in the width of $N(1535)$ in
    nuclear medium~\cite{Kim,yamazaki,yorita}.
    A more detailed calculation is reported in Ref.~\cite{Kim}.
\end{itemize}
 
Such changes in masses and $g_A$ might be observed in nuclear 
reactions with threshold or bound $\eta$ produced in a nucleus.  
For instance, the change in the mass of $N(1535)$ leads to a 
characteristic change in the potential shape of $\eta$ 
in the nucleus.  
Expected $(d, ^3He)$ cross sections are recently calculated by 
Jido, Nagahiro and Hirenzaki~\cite{jidonagahiro}.

\section{Magnetic moments in the mirror model}

In this section we discuss magnetic moments of the nucleons.  
The main interest stems from the fact that the anomalous term 
of $\sigma_{\mu \nu}$ mixes chirality just as the (Dirac) 
mass term does.  
In contrast, the normal ($\gamma_\mu$) term conserves 
chirality.  
Explicitly, the $\gamma NN$ coupling can be written as 
\beq
& & {\cal L}_{\gamma NN} = 
- \bar N \left(
\gamma_\mu Q
+ i \kappa \frac{\sigma_{\mu \nu}q^\nu}{2M} \right) 
N A^\mu \nonumber \\
&  & = \; 
- \left[ (\bar N_l \gamma_\mu Q N_l 
+ \bar N_r \gamma_\mu Q N_r )
+ i 
\left( \bar N_l \frac{\sigma_{\mu \nu}q_\nu}{2M} \kappa N_r 
+ \bar N_r \frac{\sigma_{\mu \nu}q_\nu}{2M} \kappa N_l \right)
\right]
 A^\mu\, .
\label{mugNN}
\eeq
Here $Q$ is the charge of the nucleons and 
\beq
\kappa = \kappa_S + \kappa_V \tau_3\, , 
\eeq
where 
$\kappa_S$ and $\kappa_V$ are isoscalar and isovector anomalous 
magnetic moments.  

Now let us assume the linear representation of chiral
symmetry for the nucleon again.
In the spirit of chiral symmetry, 
the electromagnetic coupling is regarded as 
a part of the chiral invariant coupling with the external 
chiral vector fields.  
In order for the anomalous coupling to be chirally symmetric, 
it should contain
the chiral field $\sigma + i \vec \tau \cdot \vec \pi \gamma_5$.  
Such a term becomes finite when chiral symmetry is broken spontaneously 
and $\bra \sigma \ket \neq 0$.  
Hence in this case, the anomalous magnetic moment should be zero
in the chirally symmetric limit.  

Another possibility is that we take the mirror model and construct
a chiral invariant term for the anomalous magnetic moment:
\beq
{\cal L}_{\rm anomalous} = 
- \frac{i}{2M}
\left(\bar N_1 \sigma_{\mu \nu} \gamma_5 \kappa N_2 
+ \bar N_2 \sigma_{\mu \nu} \gamma_5 \kappa N_1 \right) 
q^\nu A^\mu \, . 
\label{gNR_mirror}
\eeq
This is the lagrangian to the lowest order ($n = 0$) 
in powers of 
$\bra \sigma \ket^n$ and is a dominant term as
chiral symmetry is getting restored, 
$\bra \sigma \ket \to 0$.  
In the following discussion, we consider only this 
leading order term of ${\cal O}(\bra \sigma \ket^0)$, 
in order to reduce the number of free parameters.  
Hence we consider the following photon coupling term: 
\beq
{\cal L}_{\gamma NN} = 
- \left( \bar N_1 \gamma_\mu Q N_1 + 
\bar N_2 \gamma_\mu Q N_2 
+ 
(\bar N_1 \Gamma_{\mu} N_2 
+ \bar N_2 \Gamma_{\mu} N_1) q^\nu \right)
A^\mu\, , 
\eeq
where we have introduced the notation
$
\Gamma_{\mu} = (i\kappa / 2M) \sigma_{\mu \nu} q^\nu 
$.  

In terms of the physical field $N_\pm$, the coupling term
takes on the form:
\beq
{\cal L}_{\gamma NN} &=& 
- \left( \bar N_+ \gamma_\mu Q N_+ + 
\bar N_- \gamma_\mu Q N_- \right) A^\mu \nonumber \\
&-& 
\sin 2\theta 
(\bar N_{+} \Gamma_{\mu} N_{+} 
+ \bar N_{-} \Gamma_{\mu} N_{-} ) A^\mu \nonumber \\
&+& 
\cos 2\theta  (\bar N_+ \Gamma_{\mu} \gamma_5 N_- 
+  \bar N_- \Gamma_{\mu} \gamma_5 N_+) A^\mu \, .  
\eeq
From this expression, 
we find that the anomalous magnetic moments 
of $N_+$ and $N_-$ take the same value in units of 
nuclear magneton.  
Furthermore, there remain transition moments between 
the two nucleons, the $\gamma N_+ N_-$ vertex.@

Let us now briefly discuss the transition moments.  
Note that the transition term has the structure of $E1$ because of 
the parity.  
In the previous section, 
the mixing angle was estimated to be $\theta \sim 6.3 ^{\circ}$.  
We can then use the proton and neutron
magnetic moments to fix the $\kappa$'s (including the mixing angle):
$\kappa_S \sin 2\theta = -0.06$ and 
$\kappa_V \sin 2 \theta = 1.85$.  
Using these numbers, we find for the transition 
moments: 
$\mu_{pp^*} = 8.42$ and
$\mu_{nn^*} = -8.99$.  
The isovector dominance in these quantities is consistent with
experimental data, but their magnitudes are too large.  
Empirically, 
$|\mu_{pp^*}| \sim |\mu_{nn^*}| \sim 1$
in units of the nuclear magneton, as extracted from 
the helicity amplitudes~\cite{mukhopad95}:
\beq
A_{1/2}^{p} \sim 95 \times 10^{-3} {\rm GeV}^{-1/2}\, , \; \; \;  
A_{1/2}^{n} \sim -80 \times 10^{-3} {\rm GeV}^{-1/2}\, .
\eeq

In summary of this section, we have shown that 
the introduction of the anomalous magnetic
in the linear representation is different for the naive and
mirror models. 
In the naive model, the insertion of the chiral field is 
necessary in order to make the
anomalous term chirally invariant.  
Consequently,  the spontaneous breaking
gives finite values of anomalous magnetic moment.  
In contrast,  in the mirror model,
the anomalous term itself is chiral invariant.  
To the lowest order
of $\langle \sigma \rangle$ in the mirror model, we obtain $\mu_N
\sim \mu_{N^*}$ and the transition moment as a function of the mixing
angle.   
The relatively small transition moments as observed in experiment 
may suggest a larger mixing angle,
as opposed to the result obtained from 
the pion couplings previously in
eq. (3.12).  
Both facts could be an indication
that higher order terms in $\bra \sigma \ket$ could be important.
In any event, magnetic moments of the nucleon as well
as of its excited state provide useful information
of chiral symmetry of baryons.


\section{$\pi$ and $\eta$ productions at threshold region}

As discussed in the preceding sections, one of the differences 
between the naive and mirror assignments is the relative 
sign of the axial coupling constants of the positive and negative 
parity nucleons.  
Here we consider meson production reactions to observe such a sign 
difference~\cite{jido2000,jido2001}, 
by making identification once again 
$N_{+} \sim N(939) \equiv N$ and 
$N_{-} \sim N(1535) \equiv N^*$.  
From experimental point of view, $N(1535)$ has a 
nice feature that it has a strong coupling to an $\eta$, which 
can be used as a filter for the resonance production.  
In the reactions, we can observe the pion couplings which 
are related to the 
axial couplings through the Goldberger-Treiman relation
$
g_{\pi N_{\pm} N_{\pm}} {f_{\pi}} = g_{A}^{\pm} {M_{\pm}} \, .  
$

\begin{figure}
       \centerline{\includegraphics[width=12cm]
                                   {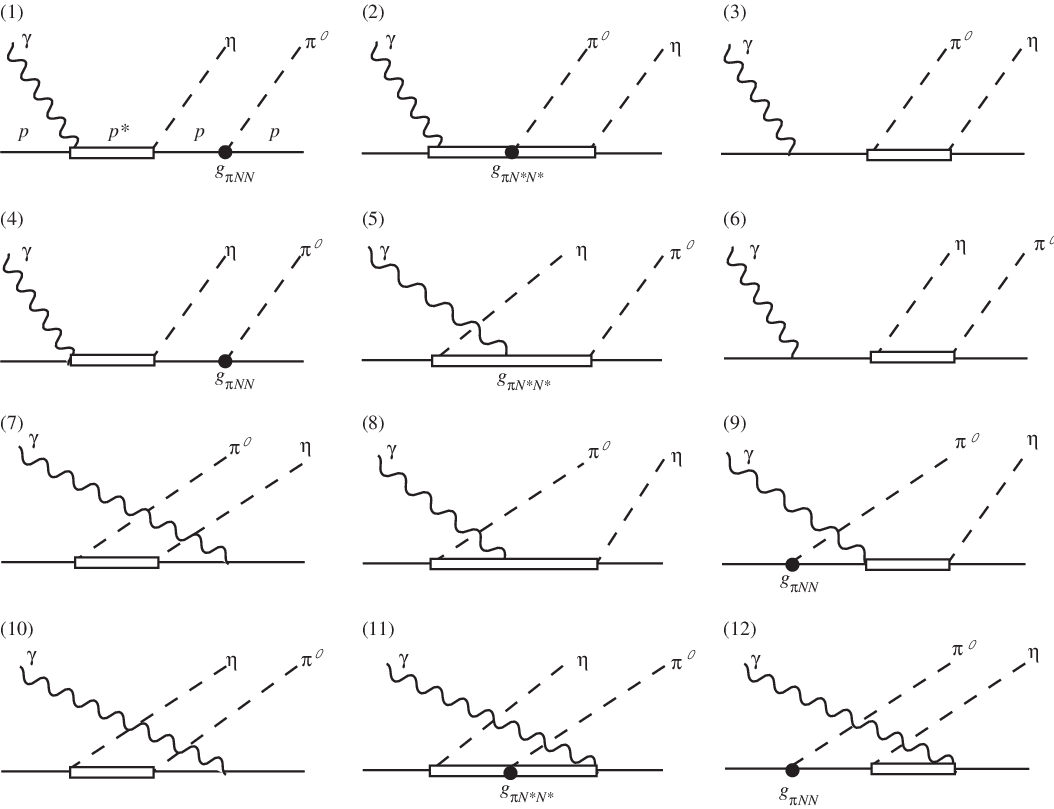}}
   \caption{Twelve diagrams for $\pi$ and $\eta$ productions.  
   The incident wavy line is either a pion of photon.  
   For $\pi^- + p \to \pi^- + \eta + p$, the first six diagrams
   contribute, while for 
   $\gamma + p \to \pi^0 + \eta + p$, 
   all twelve diagrams do.}
   \label{twelvediag}
\end{figure}

Let us consider the following two reactions:
\beq
\pi^- + p \to \pi^- + \eta + p\, , \; \; \; 
\gamma + p \to \pi^0 + \eta + p\, . 
\eeq
There are many diagrams that contribute to these reactions.  
However, suppose that the first two diagrams of Fig.~\ref{twelvediag}
are dominant.  
Then, depending on the relative sign of the $\pi NN$ and $\pi N^* N^*$ 
couplings, the two graphs are added either constructively or destructively.  
This is the essential idea to discuss the relative sign of the couplings.

In actual computation, we take phenomenological lagrangians:
\beq
L_{\pi NN}
&=& g_{\pi NN} \bar N i \gamma_{5} \vec \tau \cdot \vec \pi N \, ,
\; \; \; 
L_{\eta NN^*}
= g_{\eta NN^*} ( \bar N \eta N^*  + \bar N^* \eta N )  \, ,
\nonumber \\
L_{\pi NN^*}
&=& g_{\pi NN^*} ( \bar N \tau \cdot \pi  N^*
+ \bar N^* \tau \cdot \pi N )  \, ,
\label{Lints} \\
L_{\pi N^*N^*}
&=& g_{\pi N^* N^*} ( \bar N^* i \gamma_{5} \tau \cdot \pi  N^* )  \, .
\nonumber
\eeq
We use these interactions both for the naive and mirror cases with
empirical coupling constants for $g_{\pi NN} \sim 13$, 
$g_{\pi NN^*} \sim 0.7$
and $g_{\eta NN^*} \sim 2$.  
The coupling constants $g_{\pi NN^*} \sim 0.7$
and $g_{\eta NN^*} \sim 2$ are determined from the partial decay widths,
$\Gamma_{N^*(1535) \to \pi N} \approx  \Gamma_{N^*(1535) \to \eta N}
\sim 70$ MeV.
The unknown parameter is the $g_{\pi N^* N^*}$ coupling.
One can estimate it by using the theoretical value of the
axial coupling $g_{A}^{*}$ and the Goldberger-Treiman relation for $N^*$.
When $g_{A}^{*} = \pm 1$ for the naive and mirror assignments,
we find
$
g_{\pi N^* N^*} = g_{A}^{*} m_{N^*}/{f_{\pi}} \sim \pm 17 .
$
Here, just for simplicity, we use the same absolute value as
$g_{\pi NN}$.

The photon coupling takes on the form
\begin{equation}
    \label{LgammaNN}
  {\cal L}_{\gamma NN}= - e \bar{N} \gamma_\mu {1 + \tau_3 \over 2} N
  A^\mu + {e \over 4 M} \bar{N} (\kappa_{S} + \kappa_{V} \tau_3)
  \sigma^{\mu\nu} N F_{\mu\nu} + {\rm h.c.}\, , 
\end{equation}
where $\kappa_{S}$ and $\kappa_{V}$ are 
the physical isoscalar and isovector anomalous magnetic 
moments~\footnote{The anomalous magnetic moments here are physical ones
and are different from those in section 4},  
\begin{equation}
   \kappa_S = -0.12 \, \; \; \; \kappa_V = 3.7 \, .
\end{equation}
For $\gamma N^*N^*$, we assume the same form as (\ref{LgammaNN})
but here the nucleon mass is replaced by the resonance mass.
In this calculation, the unknown magnetic moment of $N^*$
($\equiv \kappa^{**}$) is taken to be the same as that
of the nucleon.
The uncertainty resulting from this assumption is, however,
not very important, since the diagrams
which contain the $\gamma N^* N^*$ coupling
(Figs.~\ref{twelvediag} (5) and (8)) play only a minor role.
For the $\gamma NN^*$ vertex, we employ the tensor form that is compatible
with gauge invariance:
\begin{equation}
  {\cal L}_{\gamma NN^*} =
  { i e  \kappa_{V}^* \over 2 (m_{N^*} + m_N)} \bar{N}^*
  \tau_3 \gamma_5 \sigma^{\mu\nu} N
  F_{\mu\nu} + {\rm h.c.} \, .
\end{equation}
Here we have used isovector dominance, 
and the coupling constant is given by
\begin{equation}
   \kappa_V^* = 0.9 \, ,
\end{equation}
which is determined from analyses of
eta photoproduction~\cite{mukhopad95}.

\begin{figure}
       \centerline{\includegraphics[width=13cm]
                                   {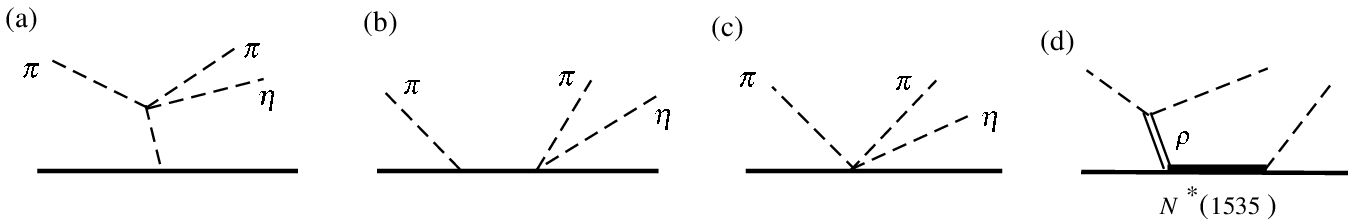}}
   \caption{Background contributions to the present reaction}  
   \label{bkground}
\end{figure}

Several remarks are in order:
\begin{enumerate}
\item 
    We have assumed resonance ($N^*$)  pole dominance.
    This is considered to be good
    particularly for the $\eta$ production at the threshold
    region, since $\eta$ is dominantly produced by $N^*$.  
\item
    For the pion induced reaction, there are several background 
	contributions as shown in Fig.~\ref{bkground} (a - d). 
	The diagrams (a - c) are suppressed due to G-parity conservation.  
	The diagram (d) is shown to be negligibly small by explicit 
	computation.  
\item
    Among various diagrams of Fig.~\ref{twelvediag}, 
    the three diagrams (1 - 3) are dominant in for the pion 
	induced process, while two (1 - 2) are important for 
	the photon induced reaction.  
\end{enumerate}

\begin{figure}
       \centerline{\includegraphics[width=13cm]
                                   {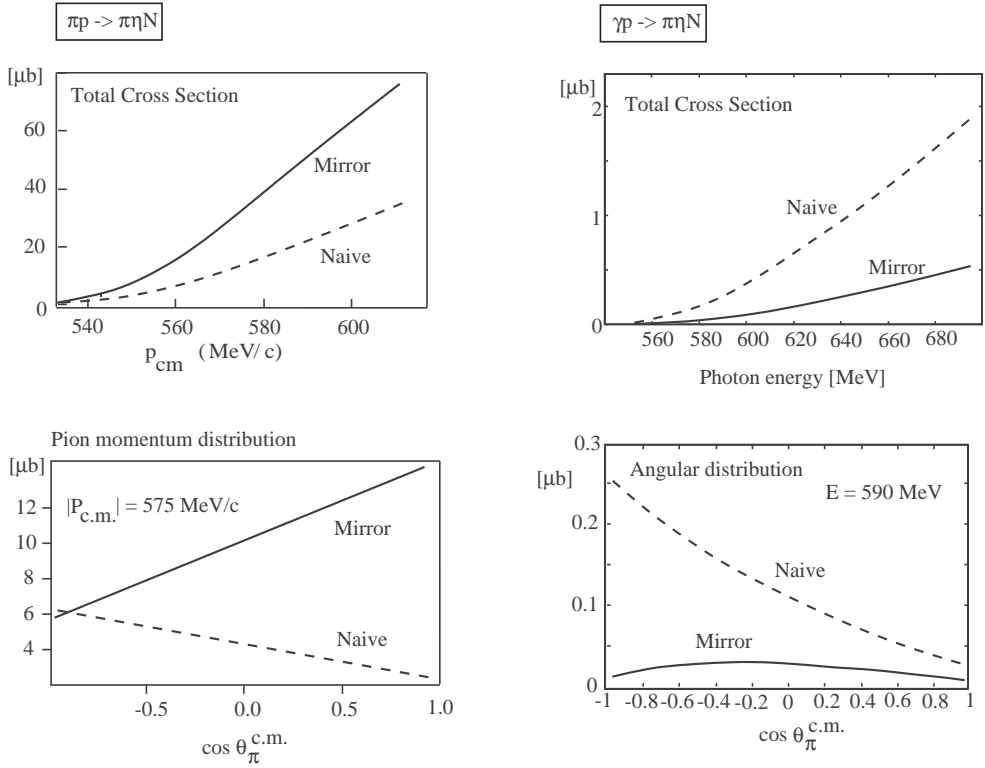}}
   \caption{Various cross sections for $\pi$ and $\eta$ productions.  
   \label{crssct}}
\end{figure}

We show various cross sections for the pion and 
photon induced processes in Fig.~\ref{crssct}.  
We observe that
\begin{enumerate}
    \item  Total cross sections are of order of micro barn, which are 
    well accessible by  experiments.  
    
    \item In the photon induced process, the two dominant 
    diagrams (1) and (2) interfere.  
	Hence for the naive model, the cross sections are enhanced, while 
	they are suppressed for the mirror model.  
    In the pion induced case, due to the momentum dependence of the initial 
    vertex the third term (3) becomes dominant as well as (1) and (2).  
        
    \item  In the pion induced reaction, the angular distribution of 
    the final state pion differs crucially.  
    They reflects the difference in the sign of the 
    $\pi NN$ and $\pi N^* N^*$ couplings.  
\end{enumerate}

\section{Summary}

In this report, we have investigated 
chiral symmetry aspects for baryon properties.  
Our investigation here is based on 
several simple linear 
representations of the chiral symmetry group.  
By assigning suitable representations, 
we can compute matrix elements of such as the axial 
coupling $g_{A}$, magnetic moments and masses, and 
put some constraints on them.   
As one of possibilities which was not considered 
before, we have studied the two different schemes 
of the naive and mirror chiral representations for the 
baryons, using linear sigma models.  
We have also proposed an experiment to distinguish 
the two chiral 
assignments in pion and eta production reactions.  
By detailed study of interfering processes, it would be 
possible to extract information on the sign of the axial coupling.  
The possibility of the negative axial coupling and the 
chiral scalar mass $m_0$ of baryons are interesting to investigate 
further.  

The description based on the linear chiral representations may be 
extended to other higher resonances, once 
we assume suitable chiral representations so as to include 
relevant resonance states.   
In Ref.~\cite{jido4}, a larger chiral multiplet 
$(1,1/2)$ was investigated.  
Since this representation contains isospin $I=1/2$ and $3/2$
component, considering
their parity partners, the four particles, 
$\Delta(1232)$, $\Delta(1700)$, $N(1720)$ and $N(1520)$
take participates in the model.  
Using the similar argument shown here, they
have obtained interesting constraints on the 
masses and decay rates of these resonances,
which are quite consistent with the observed 
properties of spin $J = 1/2, 3/2, 5/2$ sectors.
Another scheme where the nucleon $N(939)$, the delta $\Delta(1232)$ 
and the Roper $N(1440)$ are included using a larger chiral multiplet, 
$(1/2,0) + (1, 1/2)$, 
was also considered~\cite{beane1}.  
It would be interesting to 
investigate further baryon properties from the 
theory of linear chiral representations.

\end{document}